\documentclass[a4paper]{IEEEtran}
\IEEEoverridecommandlockouts
\usepackage{cite}
\usepackage{amsmath,amssymb,amsfonts}
\usepackage{algorithmic}
\usepackage{graphicx}
\usepackage{textcomp}
\usepackage[list=true, 
labelformat=brace, position=top]{subcaption}
\usepackage{xcolor}
\usepackage{mathtools}
\usepackage{multirow}
\usepackage{tikz}
\usepackage{hyperref}
\DeclarePairedDelimiter\abs{\lvert}{\rvert}
\def\BibTeX{{\rm B\kern-.05em{\sc i\kern-.025em b}\kern-.08em
    T\kern-.1667em\lower.7ex\hbox{E}\kern-.125emX}}

\def\parameters{\mathbf{\phi}}
\def\featureMap{\psi}
\usepackage[font=small]{caption}

\newcommand\QP{\mathit{QP}}

\newcommand\copyrighttext{%
	\footnotesize \textcopyright 2020 IEEE. Personal use of this material is permitted.
	Permission from IEEE must be obtained for all other uses, in any current or future 
	media, including reprinting/republishing this material for advertising or promotional 
	purposes, creating new collective works, for resale or redistribution to servers or 
	lists, or reuse of any copyrighted component of this work in other works. 
	DOI: \href{https://doi.org/10.1109/MMSP48831.2020.9287136}{10.1109/MMSP48831.2020.9287136} }
\newcommand\copyrightnotice{%
	\begin{tikzpicture}[remember picture,overlay]
		\node[anchor=north,yshift=-10pt] at (current page.north) {\fbox{\parbox{\dimexpr\textwidth-\fboxsep-\fboxrule\relax}{\copyrighttext}}};
	\end{tikzpicture}%
}
\begin{document}

%\title{Feature-Based Rate Distortion Optimization in Video Coding for Neural Object Segmentation
\title{Video Coding for Machines with Feature-Based Rate-Distortion Optimization
\thanks{The authors gratefully acknowledge that this work has been supported by the Deutsche Forschungsgemeinschaft (DFG) under contract number KA 926/10-1.}
}

\author{\IEEEauthorblockN{Kristian Fischer, Fabian Brand, Christian Herglotz, Andr\'e Kaup\\}
\IEEEauthorblockA{Multimedia Communications and Signal Processing\\}
\textit{Friedrich-Alexander-Universit\"at Erlangen-N\"urnberg (FAU)}\\
Cauerstr. 7, 91058 Erlangen, Germany \\
\{Kristian.Fischer, Fabian.Brand, Christian.Herglotz, Andre.Kaup\}@fau.de\vspace{-5mm}}

% maybe standard complient should be mentioned somewhere
% only luma channel misses as well
\maketitle
\copyrightnotice

\begin{abstract}
%\begin{itemize}
%	\item approx. 1/3 of of VVC coding gain over HEVC is achieved by the novel block partitioning \cite{wieckowski2019_pcs}
%	\item It was shown in own ICIP paper that coding gains of VVC over HEVC could not be achieved when coding for machines
%	\item Thus, developed a Rate-Distortion-Optimization that is based on the neural network feature space and therefore called FRDO
%	\item Is used to adapt block partitioning in VTM intra mode
%	\item BDR savings of up to 1.9\,\% and 4.0\,\% when allowing a delta QP of 3	
%\end{itemize}

Common state-of-the-art video codecs are optimized to deliver a low bitrate by providing a certain quality for the final human observer, which is achieved by rate-distortion optimization~(RDO). But, with the steady improvement of neural networks solving computer vision tasks, more and more multimedia data is not observed by humans anymore, but directly analyzed by neural networks. In this paper, we propose a standard-compliant feature-based RDO (FRDO) that is designed to increase the coding performance, when the decoded frame is analyzed by a neural network in a video coding for machine scenario. To that extent, we replace the pixel-based distortion metrics in conventional RDO of VTM-8.0 with distortion metrics calculated in the feature space created by the first layers of a neural network. Throughout several tests with the segmentation network Mask R-CNN and single images from the Cityscapes dataset, we compare the proposed FRDO and its hybrid version HFRDO with different distortion measures in the feature space against the conventional RDO. With HFRDO, up to 5.49\,\% bitrate can be saved compared to the VTM-8.0 implementation in terms of Bj\o ntegaard Delta Rate and using the weighted average precision as quality metric. Additionally, allowing the encoder to vary the quantization parameter results in coding gains for the proposed HFRDO of up~9.95\,\% compared to conventional VTM.

\end{abstract}

\vspace{-1mm}
\section{Introduction}

Nowadays, the amount of multimedia data that is captured to be utilized by machines instead of humans increases significantly. This machine-to-machine (M2M) communication is used for algorithms that independently solve tasks in several scopes of application, e.g., in industrial processes, detecting objects or incidents in surveillance scenarios, or in the emerging field of autonomous driving. With this increasing amount of M2M data, a suitable compression method has to be found in order to reduce the network traffic or required storage space.

\begin{figure}[t]{}
	\centering
	\includegraphics[width=0.4\textwidth]{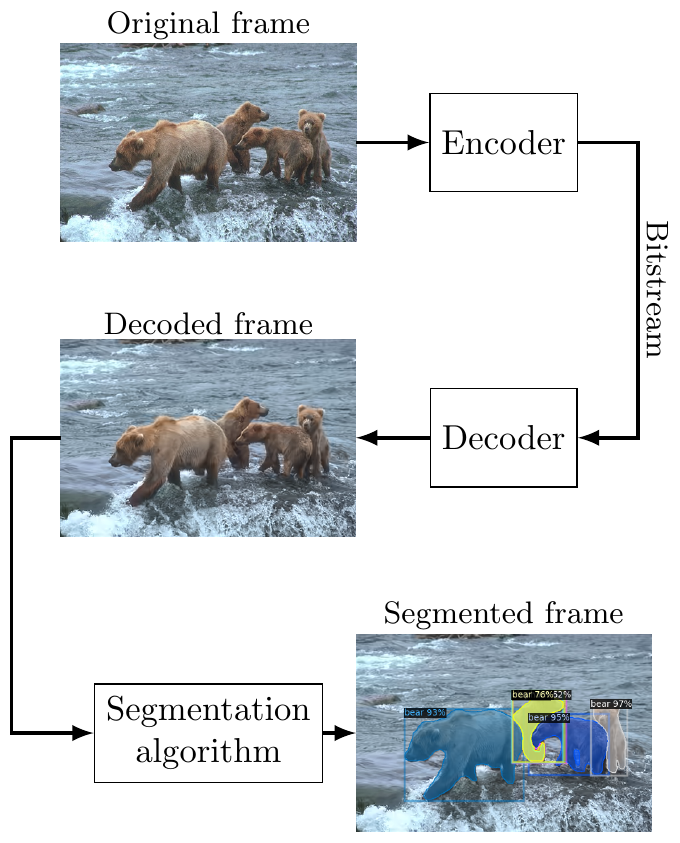}
	\caption{Exemplary VCM scenario for image segmentation.}
	\label{fig:m2m}
	\vspace{-6mm}
\end{figure}

Traditional hybrid video codecs like High-efficiency Video Coding (HEVC)\cite{sullivan2012_HEVC} and its planned successor Versatile Video Coding (VVC)\cite{bross2020_vvc}\cite{chen2020vtm8} are highly optimized to deliver the best possible quality at a specific bitrate, for which the final human user is taken as reference to evaluate the quality. Hence, the question arises, whether the optimizations made for the human user also result in the best possible compression when the compressed multimedia data is observed by an algorithm instead of a human. That question can be counted to the area of video coding for machines (VCM), for which MPEG founded an ad hoc group in 2019 targeting a possible standardized VCM bitstream format \cite{zhang2019}\cite{duan2020}. Such a VCM scenario is shown in Figure~\ref{fig:m2m} for the image segmentation task, which aims to cut out instances of objects, e.g. bears, as accurately as possible. There, the VCM task is to reduce the bitstream size while still being able to properly segment the objects in the decoded frame.

In a recent research \cite{fischer2020_ICIP}, we have discovered that the coding gains of VVC over HEVC are smaller when measuring for a VCM scenario compared to a classic scenario using quality metrics representing the human visual system. Additional experiments showed that some in-loop filters such as the adaptive loop filter, which increase the visual quality for humans, harm the coding performance when considering the detection accuracy of a neural object detection network.

Following these results, we now propose a rate-distortion optimization (RDO) for VVC that is specially designed to achieve higher coding gains in VCM scenarios compared to standard VVC encoding. To that extent, we substitute the commonly used sum of squared errors~(SSE) metric with distortion metrics that are calculated in the feature space generated with the first few layers of a neural network. We call this method feature-based RDO (FRDO), which is then taken to derive the best block partitioning decisions when the decoded frame is analyzed by an algorithm. We focused on the block partitioning since this was stated to be responsible for around a third of VVC coding gain against HEVC in~\cite{wieckowski2019_pcs}. By only influencing the encoder decisions, we assure that the resulting bitstream is still standard-compliant.

There already exist other approaches to improve the coding performance for VCM scenarios. In \cite{galteri2018}, the authors published a saliency-driven extension to the x265 codec where they reduced the quality in image areas that a saliency detection network had classified as unlikely to contain an object. A similar approach has been presented in \cite{choi2018}, where the authors used the early layers of an object detector to generate a saliency map which is used to assign fewer bits in HEVC to areas that are probably not important for object detection. A major drawback of such saliency-driven approaches is that the coding performance highly depends on the used saliency detection algorithm.

In contrast to this, our approach directly considers the influence of compression artifacts and information loss on a neural network and its convolution and pooling layers. This takes place inside the encoder and on a block-based level. Furthermore, the whole frame is encoded with the same quality and block partitioning is derived from feature space.

The remainder of this paper is structured as follows: in the subsequent chapter, the conventional RDO and the block partitioning in VVC are discussed. In Chapter~\ref{sec:frdo}, the proposed FRDO is explained in detail. The results of the conducted experiments to compare the proposed FRDO against the conventional RDO in a VCM scenario are presented in Chapter~\ref{sec:Experiments} before concluding the paper.

\section{Rate Distortion Optimization and Block Partitioning in VVC}
\subsection{Rate Distortion Optimization as Lagrangian Optimization}
For each block of an input sequence, the video encoder searches for the best possible coding parameter configuration $\parameters$ that results in a block of distorted pixels $\tilde{B}_\parameters$. To that end, the distortion $D$ has to be kept small by not exceeding a bitrate limit $R_\mathrm{max}$ which can be expressed according to \cite{sullivan1998} as 
\begin{equation}
	\min_\parameters\:D(\parameters) \quad \mathrm{s.t.} \hspace{0.5em} R(\parameters) \leq R_{\mathrm{max}},
\end{equation}
with $R(\parameters)$ representing the rate that is required to transmit the coefficients of the transformed and quantized pixel error of $\tilde{B}_\parameters$ and the metadata to signal $\parameters$ to the decoder side.

This optimization problem can be solved by a Lagrangian relaxation that balances the distortion and the rate with the Lagrange multiplier $\lambda$ by 
\begin{equation}
J(\parameters) = D(\parameters) + \lambda \cdot R(\parameters), 
\label{eq:rdo}
\end{equation}
where the encoder tests different parameter configurations $\parameters$ and chooses the configuration resulting in the lowest costs $J(\parameters)$ for a given value of $\lambda$.

Depending on the choice of $\lambda$, the RDO can be regulated towards a higher quality or a lower rate. When increasing $\lambda$, the encoder prefers a parameter configuration $\parameters$ which results in a lower rate. When decreasing $\lambda$, the encoder will find a $\parameters$ which delivers a lower distortion. The Lagrange multiplier $\lambda$ is calculated from the user-defined quantization parameter~($\QP$) according to \cite{wiegand2003} and \cite{li2013} by 
\begin{equation}
\lambda = k \cdot 2^{(\QP-12)/3},
\end{equation}
where $k$ is a constant, which is determined experimentally for each codec standard and each frame type. In the VVC test model (VTM) implementation 8.0 \cite{chen2020vtm8}, $k$ is typically set to $0.57$ for I-slices.

\begin{figure}[t]{}
	\centering
	\includegraphics[width=0.4\linewidth]{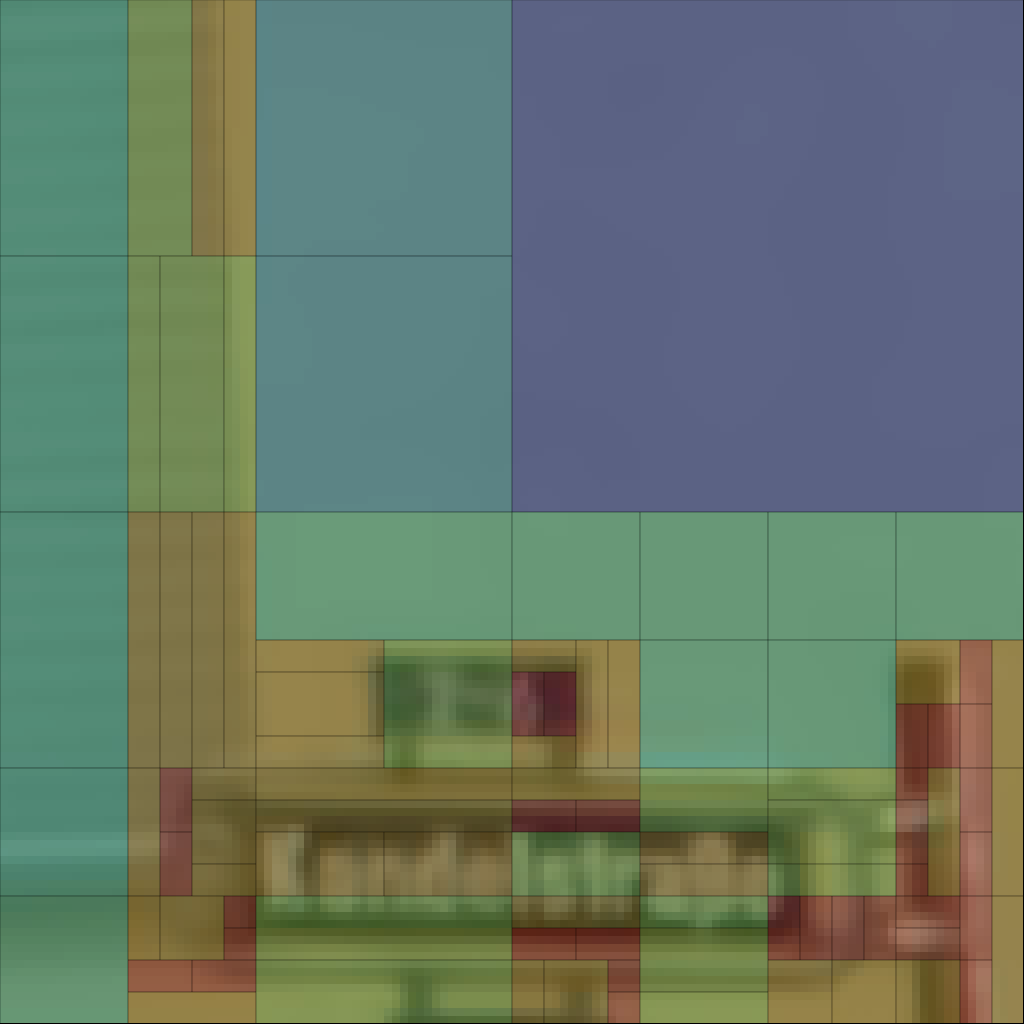}
	\caption{Block structure for an exemplary CTU of size $128\times128$ pixels, which has been coded with VTM and a $\QP$ of 22. A warmer color represents a higher depth in the recursion tree.}
	\label{fig:vtm_block_partitioning}
	\vspace{-6mm}
\end{figure}

In most hybrid video coding standards, the distortion term~$D$ for the RDO is measured with the SSE between the reconstructed pixels of a block $\tilde{B}_{\parameters}$ and the corresponding original pixels $B_\mathrm{orig}$ by
\begin{equation}
	D_{\mathrm{SSE}}(\parameters) = \sum_{(x,y)\in B}^{}(B_\mathrm{orig}[x,y] - \tilde{B}_\parameters[x,y])^2,
	\label{eq:sse_pixel}
\end{equation}
where $x$ and $y$ denote pixel indices for $\tilde{B}_\parameters$ and $B_\mathrm{orig}$.

Additionally, the rate $R$ has to be calculated for each reconstructed block in order to calculate its costs $J$. Eventually, the parameter configuration $\parameters$ resulting in the lowest costs is chosen by the encoder.

Multiple approaches already exist aiming to optimize the RDO for different use case scenarios. In \cite{wang2019_ICIP}, the authors stated that the visual impression of the human is different from the image fidelity which is measured by the SSE. Therefore, they used a hybrid approach combining the SSE distortion measurement with the structural similarity index~(SSIM) to  increase the subjective quality while still preserving the image fidelity. Another approach considers the maximum pixel error instead of the SSE which showed superior results for medical content~\cite{jaskolka_2019_ICASSP}. Lastly, the authors in~\cite{herglotz2019} proved that it is possible to decrease the decoding energy when adding an estimated decoding energy term to the RDO.

%\begin{itemize}
%	\item This optimization problem can be tackled by the Lagrangian optimization
%%	\begin{equation}
%%	\mathrm{min}\:J = D + \lambda \cdot R
%%	\label{eq:rdo}
%%	\end{equation}
%	\item in VVC, the distortion is measured with the SSE.
%	\item SSE for a block of pixels $B$ is calculated by (with $X_\mathrm{orig}$ being the block with original pixels and $\tilde{X}_\mathrm{rec}$ the block that has been reconstructed from the predicted and quantized coefficients)
%	\begin{equation}
%		D_{\mathrm{SSE}}(B) = \sum_{(x,y)\in B}^{}(X_\mathrm{orig}[x,y] - \tilde{X}_\mathrm{rec}[x,y])^2
%	\end{equation}
%	\item rate is also estimated for each block
%	\item these costs have to be compared for each coding block resulting from a different parameter set
%	\item is used to find the best intra mode and the best transformation for each block with the lowest costs $J$
%\end{itemize}

\begin{figure*}
	\centering
	\begin{subfigure}[t]{0.49\textwidth}
		\centering
		\includegraphics[width=\textwidth]{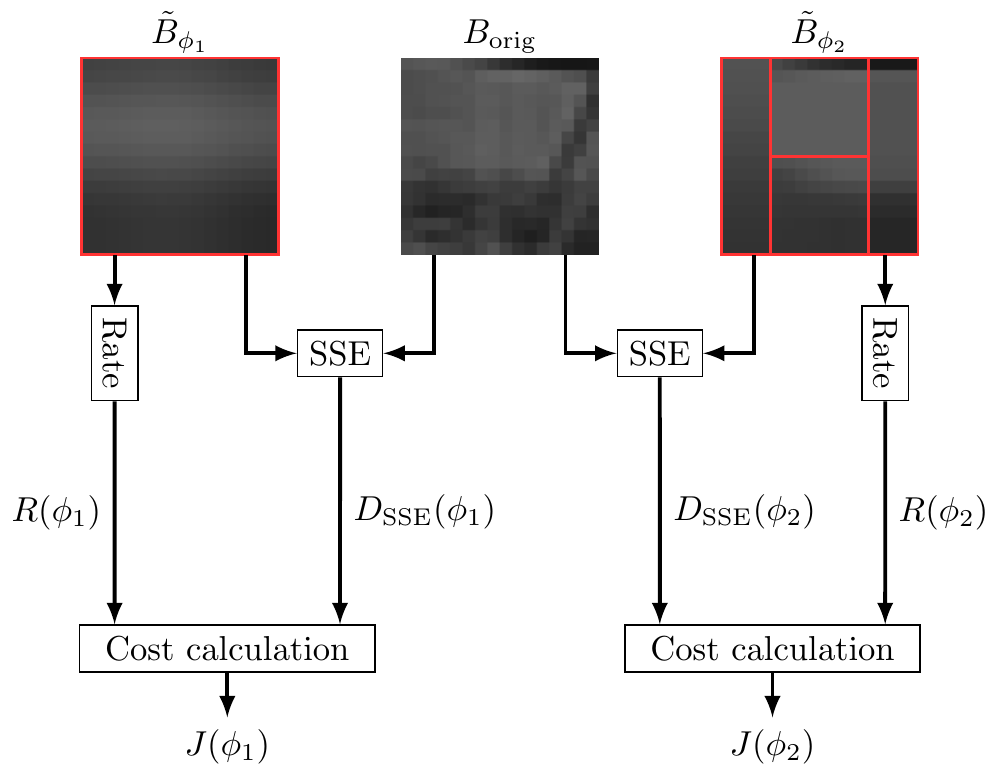}
		\subcaption{Classic RDO}
		\vspace{-4mm}
		\label{fig:rdo}
		
	\end{subfigure}
	~
	\begin{subfigure}[t]{0.49\textwidth}
		\centering
		\includegraphics[width=\textwidth]{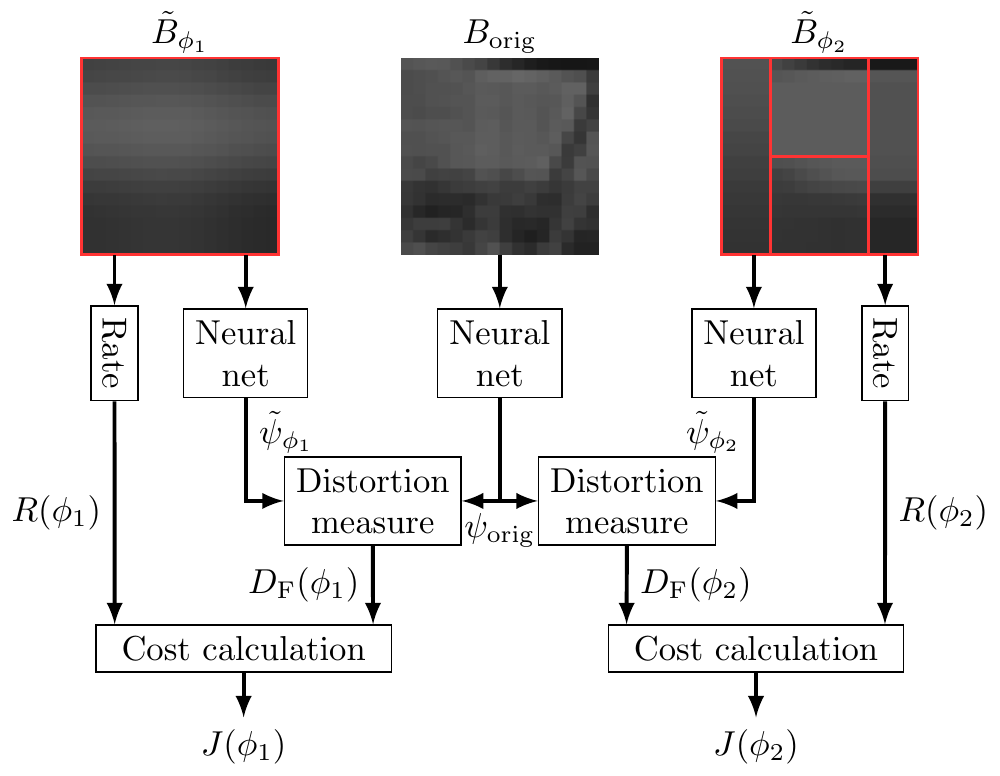}
		\subcaption{Proposed FRDO}
		\vspace{-2mm}
		\label{fig:frdo}
		
	\end{subfigure}
	
	\caption{Comparison between RDO with FRDO for two reconstructed blocks $\tilde{B}$ from original block $B_\mathrm{orig}$, resulting from two different parameter configurations $\parameters_1$ and $\parameters_2$. The red lines indicate the block partitioning.}
	\label{fig:comparison rdo frdo}
	\vspace{-6mm}
\end{figure*}

\subsection{Block Partitioning in VVC with RDO}
\label{subsec: Block partitioning in VVC with RDO}
In VVC, each frame is separated into coding tree units~(CTUs) with a size of $128\times128$ pixels \cite{chen2020vtm8}. These CTUs are used as the root of a quad-tree, which equally separates the CTUs recursively into four squared coding units (CUs). Thereupon, each leaf of the quad-tree can be further divided into non-squared CUs by a nested multi-type tree, which is new compared to HEVC. The multi-type tree allows binary and ternary splits in horizontal and vertical direction so that the encoder is able to adapt the CUs more precisely to the frame content. This recursive procedure can split the CUs up to a size of $4\times4$ pixels. An exemplary VVC block partitioning for a CTU is depicted in Figure \ref{fig:vtm_block_partitioning}, where it can be seen that homogeneous areas are covered with large CU sizes, whereas the encoder chooses smaller CUs in areas with more detailed content such as the street sign in the bottom part. The binary and ternary splits result in non-squared CUs, which allow the encoder more freedom to adapt the block partitioning to the content.

In order to achieve the best possible CTU partitioning in terms of RDO, each CU of the size $M\times N$ is compared against a CU of the same size, which has been split differently into sub-CUs as it can be seen for the corresponding coding blocks $\tilde{B}_{\parameters}$ in Figure~\ref{fig:rdo}. For both CUs the best parameter configuration, which contains the prediction and the transform mode, has already been determined before, and thus the distortion and the rate including the additional bits for signaling the splits can be calculated. Then, the CU with the lower costs will be used for further processing in the coding tree at larger CU sizes. This procedure starts at the deepest recursion level leaves with small CU sizes, and moves up to the quad-tree root until the complete CTU is partitioned.

%\begin{itemize}
%	\item As stated in the VTM description \cite{chen2020vtm8}, to process each frame, each intra frame is separated into CTUs with the size of 128x128 pixels. These CTUs are subdivided using a recursive quadtree partitioning into coding units (CU). Compared to HEVC, now, a multi-type tree providing binary and ternary splits is added. Thus, the coding structure can be adapted better to the frame content. Example is shown in Figure ...
%	\item Now CU is used for prediction and transform
%	\item Each CTU is used as root of the quad-tree and each leaf of the quad-tree can be further divided into non-squared CUs by an adjacent multi-type tree.
%	\item smallest size is 4x4
%	\item in order to get best the best coding tree so far, the reconstructed pixels of the current block $R_\mathrm{cur}$ are compared against the best 
%	
%\end{itemize}

\section{Proposed Feature-Based RDO}
\label{sec:frdo}
\subsection{Introduction to FRDO}
Most neural networks take the input information from pixel space and transform this information into a denser and more abstract feature space by convolution and pooling layers. From this compact feature space, the neural network derives decisions with the help of attached fully-connected and softmax layers, e.g. to locate or classify objects. Therefore, the task of VCM is to preserve a high feature fidelity rather than a high pixel fidelity. However, while a high pixel fidelity automatically leads to a high feature-fidelity, the reverse is not the case, since a high feature fidelity does not necessarily require a high pixel fidelity. The proposed FRDO aims at exploiting this gap in order to reduce the required rate.

As shown in Figure~\ref{fig:frdo}, the first step of FRDO transforms each coding block $B_\parameters$ and the corresponding original block $B_\mathrm{orig}$ into the feature space. For that purpose, the first convolution, Rectified Linear Unit~(ReLU), and pooling layers of an arbitrary trained classification or detection network are fed with the block content, which can be formulated as  
\begin{equation}
\featureMap_\parameters = f(B_\parameters), 
\end{equation}
with $f(\cdot)$ and $\featureMap_\parameters$ representing the used neural network and the resulting three-dimensional feature map, respectively. The reasons for only using the first layers of a neural network to derive the feature space are twofold. First, the runtime is reduced. Second, the spatial resolution is decreased by the subsequent convolution and pooling layers, which makes deep networks unfeasible to apply to small coding blocks. Even by zero-padding small input patches, e.g. $4\times4$ pixels in VVC, these padded patches then consist of a significant number of zeros which is likely to falsify the distortion measurement.

%As shown in Figure~\ref{fig:frdo}, the proposed FRDO replaces the SSE-based distortion metric $D_{\mathrm{SSE}}$ with a distortion metric $D_\mathrm{F}$ that is based on the feature map $\featureMap$ gained by applying the first convolution and pooling layers of an arbitrary classification or detection network. This cropped network is applied on a coding block $B$ with the size of $M\times N \times 1$ pixels, since one channel is encoded after another in the encoder. However, the common classification and detection networks are designed for a three-channel input image. Thus, we concatenate $B$ three times with itself to get a patch of size $M\times N \times 3$, which we call $B_\mathrm{concat}$ from now on, before applying the neural network by
%
% Due to zero-padding, the resulting feature map $\featureMap$ has the same spatial dimensions as the input block, but the number of channels $C$ increases for common neural networks.

\subsection{Calculating Distortions in Feature Space}

Now, the feature map $\tilde{\featureMap}_\parameters$ derived from the reconstructed block $\tilde{B}_\parameters$ can be compared against the feature map $\featureMap_\mathrm{orig}$ derived from the original block $B_\mathrm{orig}$. With that, a distortion can be measured in this feature space similar to \eqref{eq:sse_pixel} by
\begin{equation}
	D_{\mathrm{FSSE}}(\parameters) = \sum_{(x,y,c) \in \featureMap}^{}(\featureMap_\mathrm{orig}(x,y,c) - \tilde{\featureMap}_\parameters(x,y,c))^2,
\end{equation}
with $x$, $y$, and the channel index $c$ representing the pixel-wise access to each element of $\featureMap_\mathrm{orig}$ and $\featureMap_\parameters$ in the feature space.

Finally, this new feature-based distortion measure $D_{\mathrm{FSSE}}$ allows to reformulate the RDO cost calculation from \eqref{eq:rdo} as FRDO for each parameter configuration $\parameters$ as
\begin{equation}
\min_\parameters\:J(\parameters) = D_{\mathrm{FSSE}}(\parameters) + \lambda \cdot R(\parameters).
\label{eq:frdo}
\end{equation}

As generally known, SSE tends to focus on large errors by neglecting smaller ones because of the square operation. Thus, we also propose a feature-based sum of absolute differences~(FSAD), which can be formulated as
\begin{equation}
D_{\mathrm{FSAD}}(\parameters) = \sum_{(x,y,c) \in \featureMap}^{}\abs{\featureMap_\mathrm{orig}(x,y,c) - \tilde{\featureMap}_\parameters(x,y,c)}.
\end{equation}

A comparison between the feature-based distortion metrics is provided in Section \ref{sec:Experiments}.
%\begin{itemize}
%	\item Proposed FRDO replaces $D_{\mathrm{SSE}}$ by a distortion metric that is based on the features $\featureMap$ gained from applying the first 3 layers of a VGG-16 network \cite{simonyan15} on a block of grayscale pixels $X$ 
%
%	\item 
%	\item The feature-based distortion metric $D_\mathrm{FB}$ is now calculated by comparing the feature maps of two different coding blocks with the corresponding original block
%	\item New distortion metric SSE is calculated in feature space for a block B:
%	\begin{equation}
%		D_{\mathrm{FB, SSE}}(B) = \sum_{x,y,c \in \Theta}^{}(\featureMap{B,\mathrm{orig}}(x,y,c) - \tilde{\featureMap}{B, \mathrm{rec}}(x,y,c))^2
%	\end{equation}
%	\item in this case $\mathbf{z}$ represents each element in the three-dimensional feature space
%	\item with this new distortion measure, the cost calculation from \eqref{eq:rdo} can be reformulated for each block B as
%	\begin{equation}
%		\mathrm{min}\:J(B) = D_{\mathrm{FB, SSE}}(B) + \lambda \cdot R(B)
%		\label{eq:frdo}
%	\end{equation}
%\end{itemize}

\subsection{Adapt Feature-Based Distortion Measure to $\lambda$}
%However, the values resulting from a feature-based distortion $D_\mathrm{F}$ do not necessarily have to be located in the same range as for the common distortion measure $D_\mathrm{SSE}$. Nevertheless, this is highly required to preserve the balance between the distortion and the cost term, since the general $\lambda$ values have been derived for SSE as distortion measure.
As shown in \eqref{eq:rdo} and \eqref{eq:frdo}, RDO and FRDO both depend on a carefully chosen Lagrange multiplier $\lambda$. The $\lambda$ used in the VTM encoder is optimized for SSE. However, the proposed feature-based distortion $D_\mathrm{F}$ can be located in a different range of values than $D_\mathrm{SSE}$, which might lead to sub-optimal coding decisions. To relieve this problem, we scale the feature-based distortion derived from the first parameter configuration $\parameters_1$ to the equivalent SSE by
\begin{equation}
	D^{*}_{\mathrm{F}}(\parameters_1) = D_{\mathrm{F}}(\parameters_1) \cdot \frac{D_{\mathrm{SSE}}(\parameters_1)}{D_{\mathrm{F}}(\parameters_1)} = D_{\mathrm{SSE}}(\parameters_1),
\end{equation}
which results in the feature-based distortion and the SSE for a block  coded with $\parameters_1$ having the same value. The feature-based distortion of the second coding block is multiplied with the same factor, in order to preserve the relative differences by
\begin{equation}
	D^{*}_{\mathrm{F}}(\parameters_2) = D_{\mathrm{F}}(\parameters_2) \cdot \frac{D_{\mathrm{SSE}}(\parameters_1)}{D_{\mathrm{F}}(\parameters_1)}.
\end{equation}

With this mapping, the feature-based distortion is adapted to $\lambda$ by still being able to compare the two coding blocks according to their influence on a neural network, without having to re-evaluate $\lambda$ for the proposed distortion measures.

With this normalized distortion and since it was shown in~\cite{wang2019_ICIP} and~\cite{herglotz2019} that it is beneficial to not completely omit the common SSE as distortion, we also propose a hybrid FRDO~(HFRDO) version that equally combines $D_\mathrm{SSE}$ and a normalized feature-based distortion $D^*_\mathrm{F}$ by
\begin{equation}
J(\parameters) = 0.5 \cdot(D_\mathrm{SSE}(\parameters) + D^*_\mathrm{F}(\parameters)) + \lambda \cdot R(\parameters).
\label{eq:hybrid}
\end{equation}

\subsection{Applying FRDO in Hybrid Video Codec}
FRDO can substitute every RDO in common hybrid video codecs where coding parameter configurations $\parameters$ and thus the corresponding coding blocks are compared to each other. In this paper, we focus on the block partitioning as explained in Section~\ref{subsec: Block partitioning in VVC with RDO} to save complexity.

In further experiments, we also allow the encoder to vary the $\QP$ for each CU in a certain range, by using the FRDO. There, the encoder partitions a CTU into the blocks for different $\QP$ values, and chooses the best $\QP$ in terms of rate and distortion afterwards. With that, the coding decisions can be even better adapted to neural networks at the drawback of increased encoding complexity depending on the allowed $\QP$ range.

\begin{figure}[t]{}
	\centering
	\includegraphics[width=0.49\textwidth]{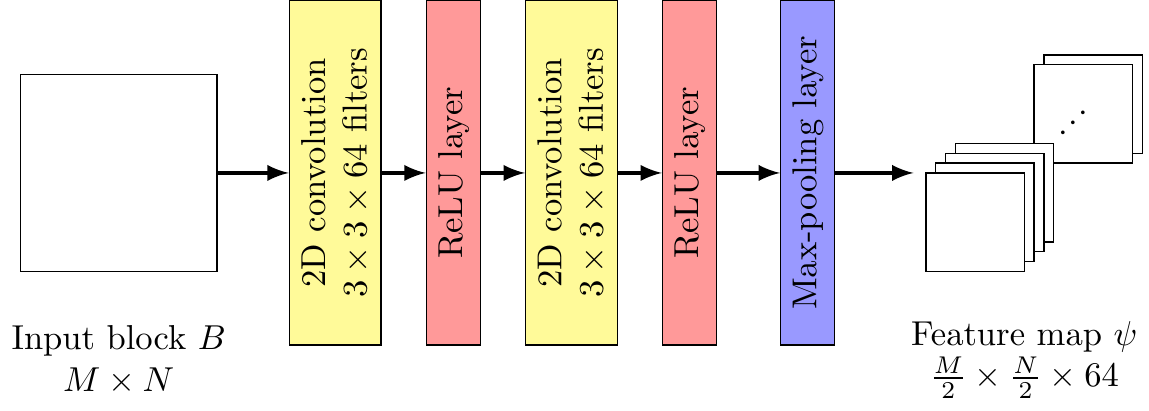}
	\caption{Extracting the feature map $\featureMap$ from a block of pixels $B$ with the first five VGG-16 layers.}
	\label{fig:vgg}
	\vspace{-4mm}
\end{figure}

\section{Experimental Results}
\label{sec:Experiments}

\subsection{Simulation Setup}

In order to evaluate the proposed FRDO, we added it to VTM 8.0~\cite{chen2020vtm8} and compared this FRDO VTM against the unmodified version with conventional RDO. We only applied the FRDO to the luminance channel. To derive the feature space for FRDO, we used a pre-trained VGG-16 \cite{simonyan15} model from the Pytorch vision package in version 0.5.0~\cite{pytorch2020} that is actually used to classify images. From this model, we used the first five layers as shown in Figure \ref{fig:vgg}. In order to measure the coding performance for neural networks, we compressed the 500 validation images of the Cityscapes~\cite{cordts2016} dataset, which contains street views with annotated road users, with $\QP$ values in all intra configuration. Subsequently, each compressed frame is fed into Mask R-CNN~\cite{he2017} to detect and extract all road users. The extracted objects are compared against the ground-truth instances with the mean average precision~(mAP) metric as described in \cite{cordts2016} and implemented in~\cite{cordts2017cityscapesscripts} that directly measures the segmentation performance of Mask R-CNN. Similar to~\cite{fischer2020_ICIP}, we weight the average precision~(AP) according to the number of instances of each classes in the validation set, since some classes such as bus and truck are underrepresented in Cityscapes compared to the classes car and person. Because of that, weighting all classes equally would lead to a deformed result. For Mask R-CNN, we chose the model with a ResNet50 backbone from the Detectron2 model zoo \cite{wu2019detectron2}, which has already been trained on the Cityscapes dataset. By using a validation network that differs in structure and weights from the VGG-16 network to obtain the feature space, we want to show the universal applicability of FRDO without any knowledge of the finally used network in the application case.

\subsection{Comparison FRDO with RDO}

\begin{figure}[t]{}
	\centering
	\includegraphics[width=0.49\textwidth]{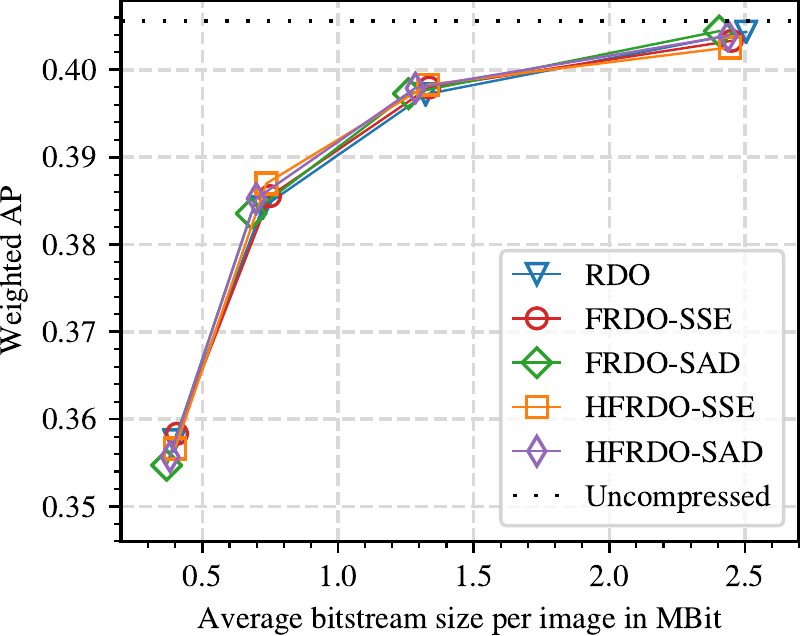}
	\caption{Weighted AP values over bitrates comparing the different RDOs. Values are averaged over the 500 Cityscapes validation images for $\QP \in \{12, 17, 22, 27\}$.}
	\label{fig:map-rate curve}
%	\vspace{-4mm}
\end{figure}

\def\imSize{3cm}
\begin{figure}[t]{}
	\centering
	
	\begin{subfigure}[t]{\imSize}
		\centering
		\includegraphics[width=\textwidth]{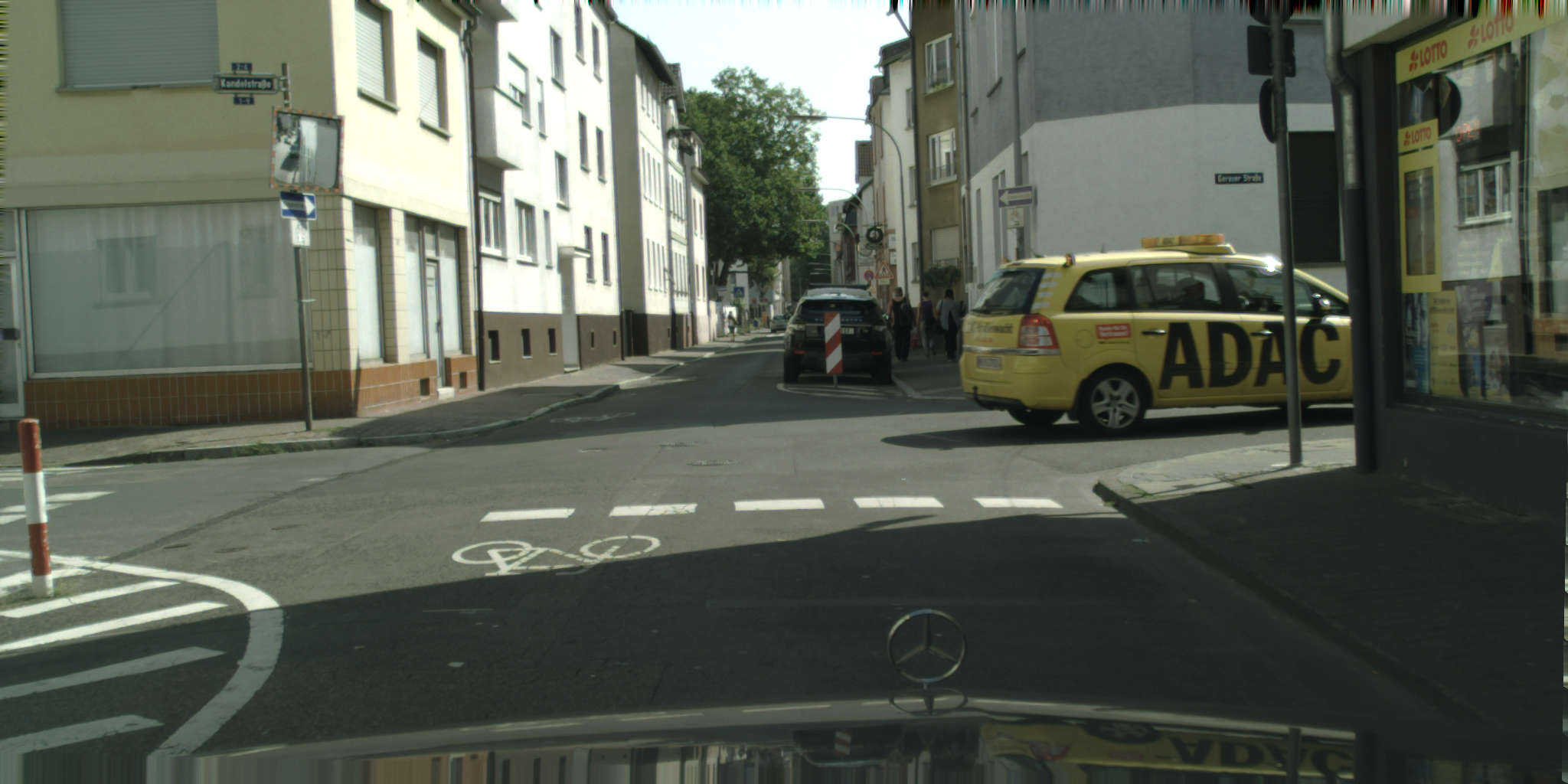}
		\subcaption{Original Frame}
	\end{subfigure}
	\hfil
	\begin{subfigure}[t]{\imSize}
		\centering
		\includegraphics[width=\textwidth]{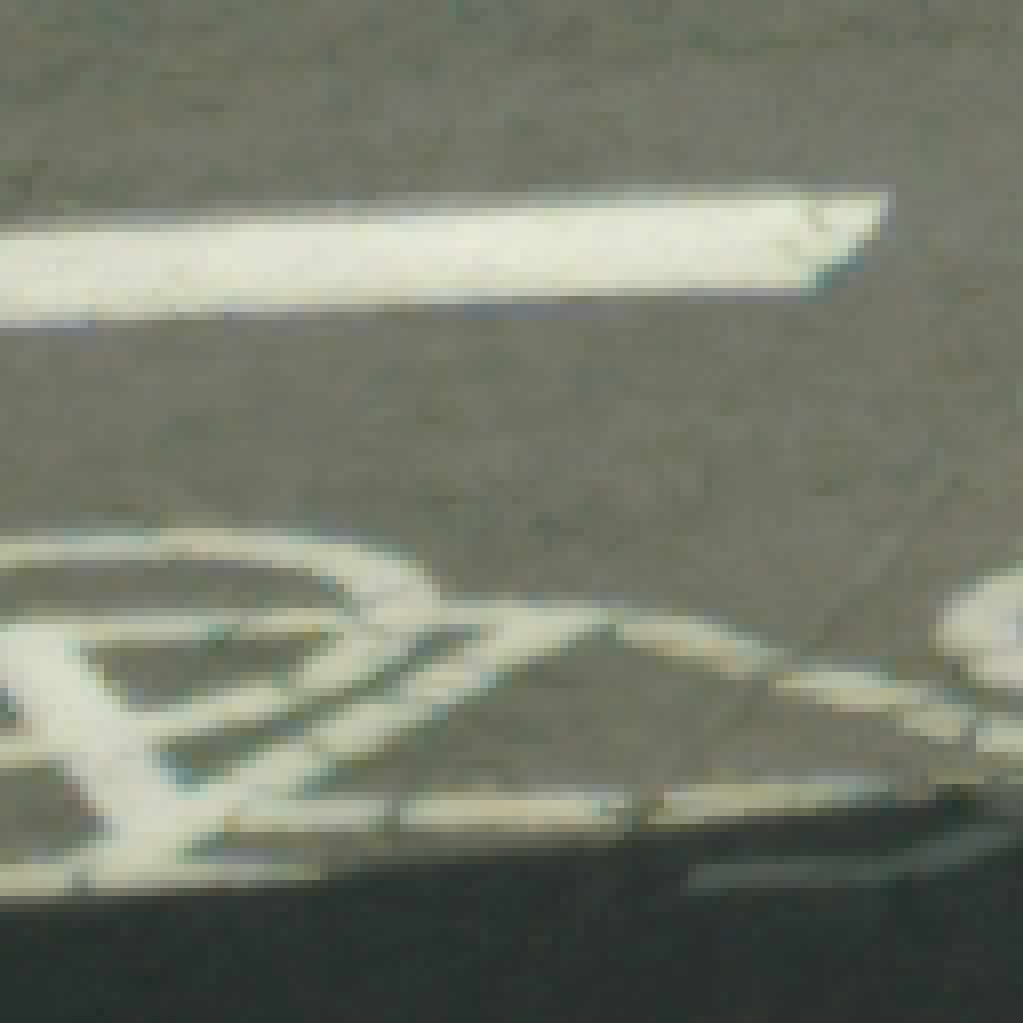}
		\subcaption{Original CTU}
	\end{subfigure}

	\begin{subfigure}[t]{\imSize}
		\centering
		\begin{tikzpicture}[]
		    \draw (0,0) node[inner sep=0] {\includegraphics[width=\textwidth]{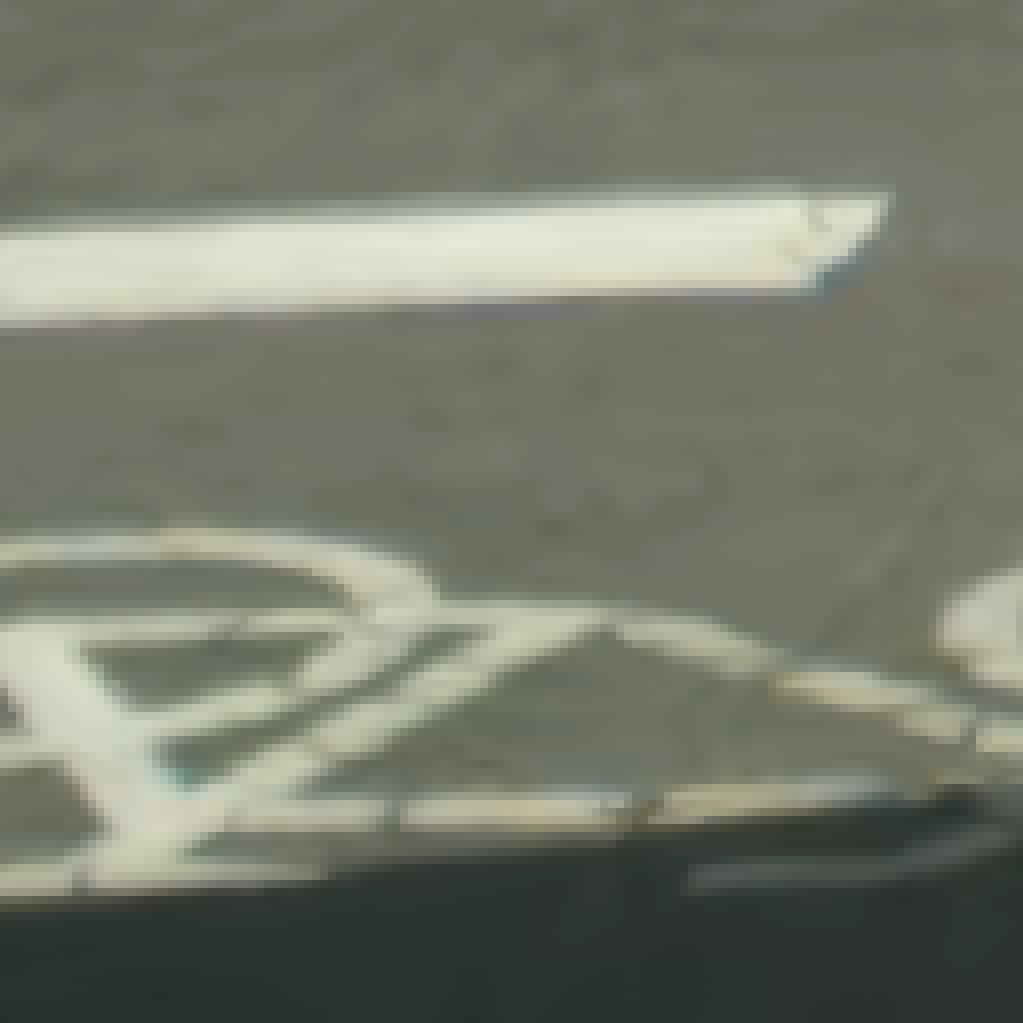}};
		    \node(text) [rectangle, fill=white, anchor=north west, inner sep=2pt] at (-\textwidth/2,\textwidth/2) {\scriptsize Y-PSNR=44.4\,dB};
		\end{tikzpicture}
		\vspace{-3mm}
		\subcaption{CTU coded with RDO}
	\end{subfigure}
	\hfil
	\begin{subfigure}[t]{\imSize}
		\centering
		\begin{tikzpicture}[]
			\draw (0,0) node[inner sep=0] {\includegraphics[width=\textwidth]{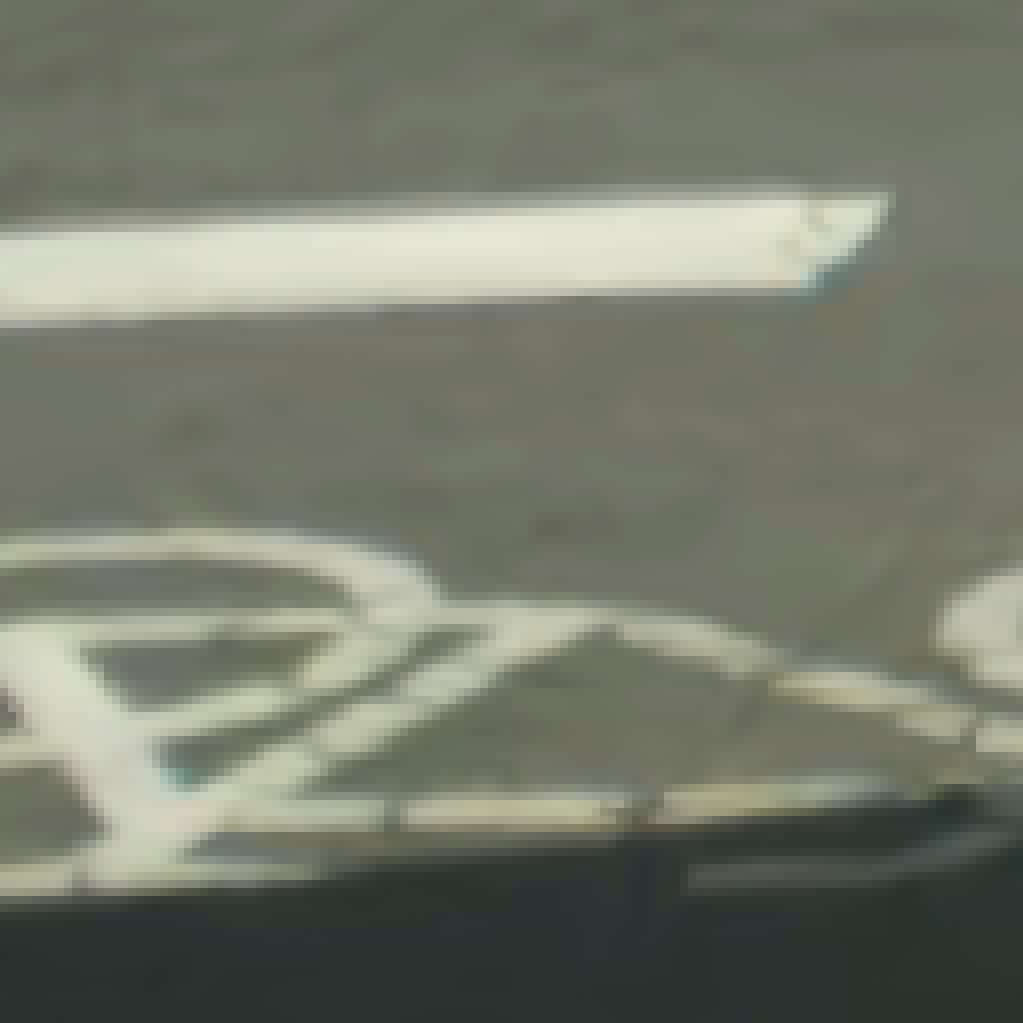}};
			\node(text) [rectangle, fill=white, anchor=north west, inner sep=2pt] at (-\textwidth/2,\textwidth/2) {\scriptsize Y-PSNR=43.8\,dB};
		\end{tikzpicture}
		\vspace{-3mm}
		\subcaption{CTU coded with FRDO}
	\end{subfigure}
	
	\begin{subfigure}[t]{\imSize}
		\centering
		\includegraphics[width=\textwidth]{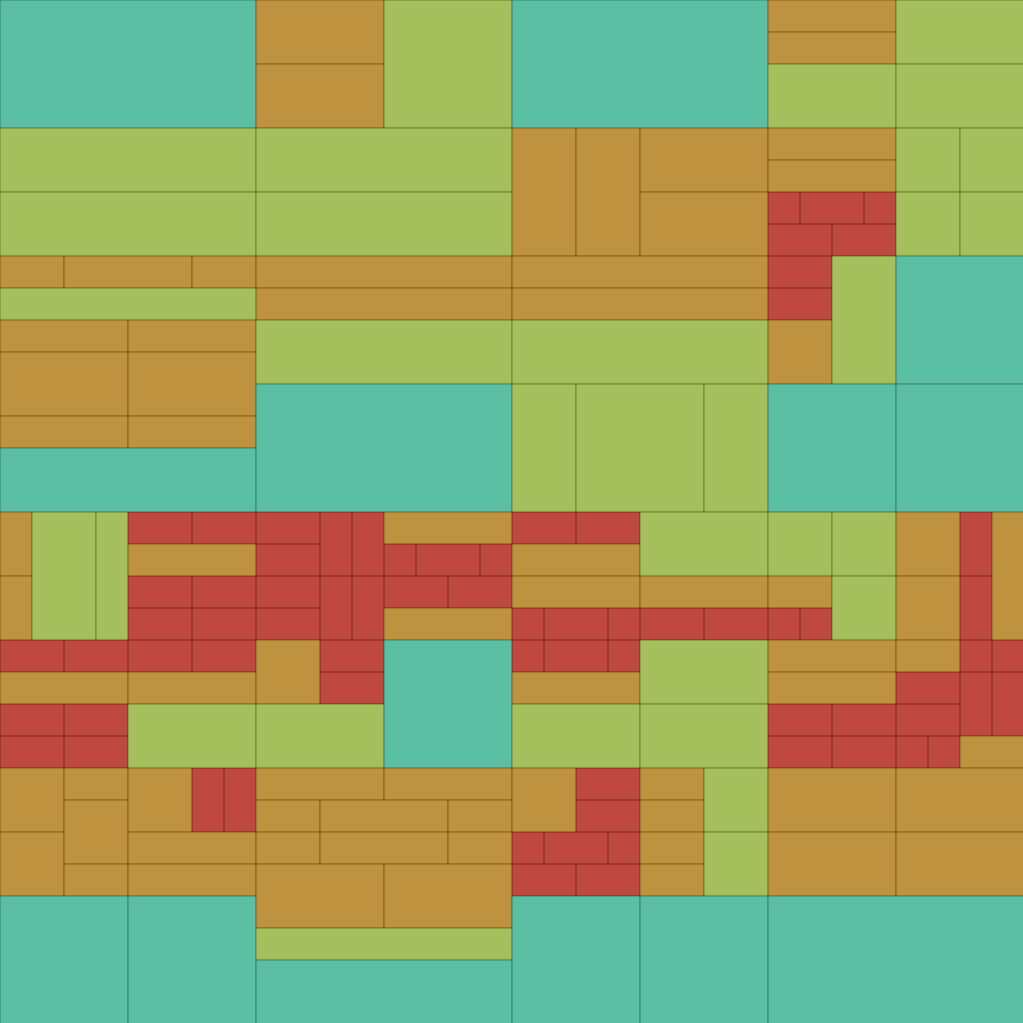}
		\vspace{-3mm}
		\subcaption{Block structure RDO \\}
	\end{subfigure}
	\hfil
	\begin{subfigure}[t]{\imSize}
		\centering
		\includegraphics[width=\textwidth]{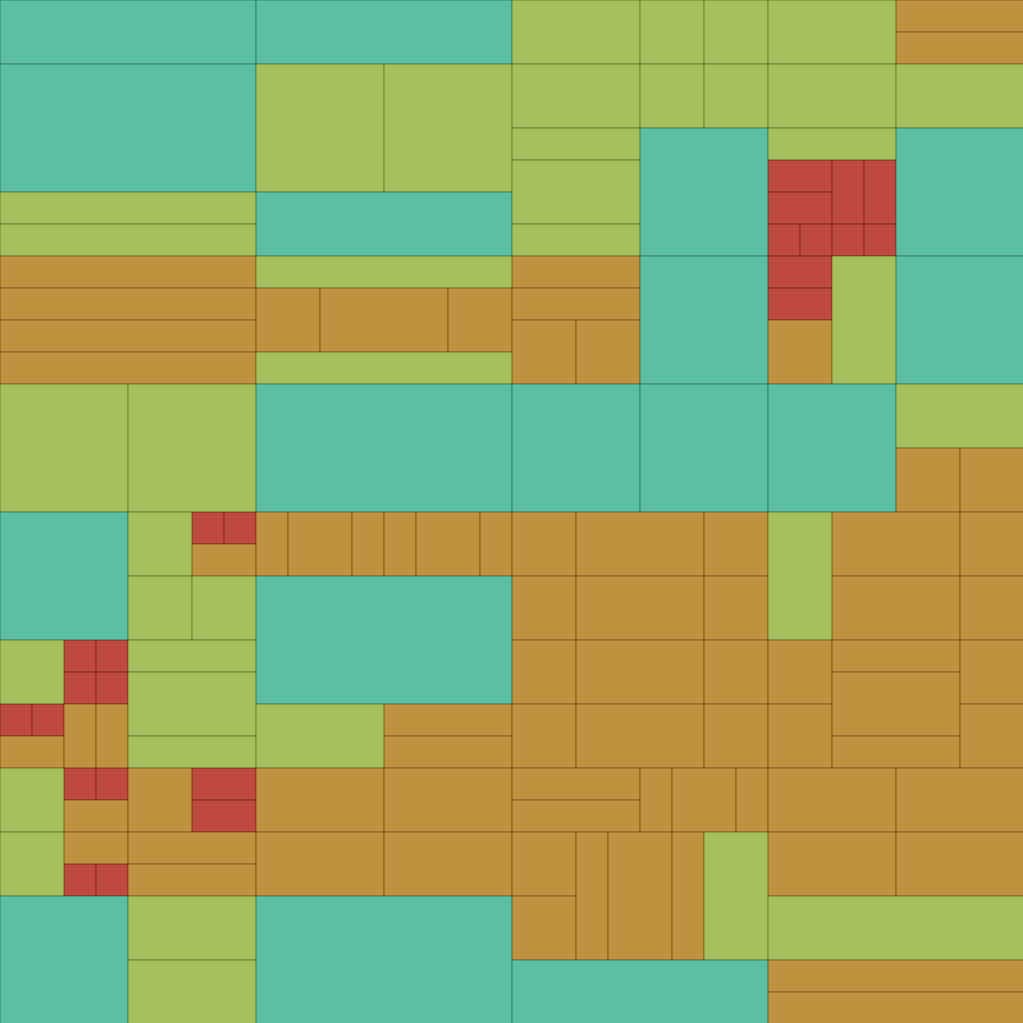}
		\vspace{-3mm}
		\subcaption{Block structure FRDO}
	\end{subfigure}
	
	\caption{Comparison of one exemplary CTU out of Cityscapes image \textit{frankfurt\_000000\_000294\_leftImg8bit.png} compressed with RDO (left) and FRDO with SAD (right) and a static $\QP$ of 22. A warmer color represents a higher depth in the recursion tree.}
	\label{fig:visual comparison}
	\vspace{-5mm}
\end{figure}

In Figure~\ref{fig:map-rate curve}, the weighted AP for Mask R-CNN is plotted over the required bitrate for conventional RDO as reference and FRDO with the different distortion metrics. We selected a $\QP$ range from 12 to 27 in steps of 5, because for most practical applications the accuracy loss caused by compression should not be too high compared to the case with uncompressed input data (dotted line).

It can be seen from the red curve that FRDO with SSE as distortion measure in the feature space is not able to significantly reduce the required bitrate compared to the blue curve representing the standard VTM with normal RDO. The green curve representing the proposed FRDO with SAD requires less bitrate than RDO, while maintaining a similar detection rate for high bitrates. With decreasing bitrate, the FRDO with SAD is not able to keep the same detection rate as the reference. The resulting bitrate for the measurement points of the hybrid FRDO as introduced in \eqref{eq:hybrid} with SAD are located, as expected, between the points for RDO and FRDO-SAD.

\begin{table}[t]
	\normalsize
	\caption{BDR with respect to the particular quality metric using VTM with conventional RDO as anchor for $\QP\in\{12, 17, 22, 27\}$.}
	\label{tab: BDR QP}
	\centering
	\begin{tabular}{lrr}
		\hline
		                    & \multicolumn{2}{c}{Quality metric} \\
Optimization method &  PSNR               & Weighted AP \\  \hline
		FRDO-SSE         	&4.30\,\%  &-0.50\,\%     \\
		FRDO-SAD           &3.02\,\%    &-3.99\,\%   \\
		HFRDO-SSE           &1.63\,\%    &-4.55\,\%   \\
		HFRDO-SAD           &1.26\,\%    &-5.49\,\%  \\
		\hline
	\end{tabular}
	\vspace{-4mm}
\end{table}

In order to quantify the coding performance of the different FRDO versions, the Bj\o ntegaard delta rate~(BDR)~\cite{bjontegaard2001} values are listed in Table \ref{tab: BDR QP} for the curves shown in Figure~\ref{fig:map-rate curve}. The BDR values with respect to the PSNR represent the bitrate savings at a fixed quality for the human visual system. Calculating the BDR with respect to the weighted AP returns the bitrate savings at a fixed detection rate for Mask R-CNN. Thereby, a negative number expresses savings compared to the anchor. 

From this table, it can be seen that the FRDO with SSE reduces the required bitrate by 0.5\,\% for the same detection accuracy. Using FRDO-SAD decreases the bitrate by 3.99\,\%. These results indicate that SAD is better suited to represent distortions in the feature space than SSE. At the same time, the coding performance measured with PSNR decreases for the FRDO variants, which proves that the proposed FRDO can preserve the feature space even though it significantly reduces the pixel-fidelity. Using the hybrid FRDO with SSE achieves coding gains of 4.55\,\% compared to the standard RDO in VTM. Using the HFRDO-SAD results in the best coding performance for the VCM task with 5.49\,\% bitrate savings. Compared to FRDO, the hybrid variants show higher coding gains for mAP and do not harm the classic coding performance for PSNR that much. 
%All in all, it is shown that the proposed FRDO and HFRDO result in coding gains for VCM scenarios even though the layers to determine $D_\mathrm{F}$ differ from the validation network. This indicates that the proposed FRDO is universally applicable for arbitrary networks.

In Figure~\ref{fig:visual comparison}, the block partitioning with RDO and FRDO with SAD is compared for a CTU that does not contain a relevant object. With FRDO, the block structure is not as deep as for the same CTU coded with RDO, which results in bitrate savings. Yet, the pixel-fidelity and also the visual quality is slightly better for the CTU compressed with RDO.

\subsection{FRDO with Delta $\QP$}
\begin{figure}[t]{}
	\centering
	\includegraphics[width=0.49\textwidth]{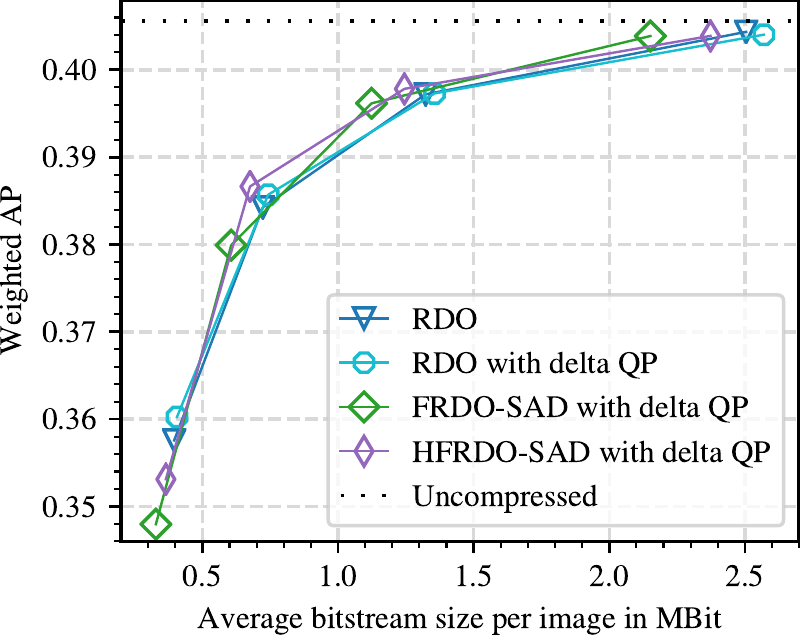}
	\caption{Weighted AP values over bitrates comparing the different RDOs when allowing delta $\QP$. Values are averaged over the 500 Cityscapes validation images for $\QP \in \{12, 17, 22, 27\}$.}
	\label{fig:map-rate curve delta qp}
\end{figure}

In another experiment, we also allowed the encoder to vary the base $\QP$ in a range of $-3$ to $+3$ for each CU according to the used RDO and plotted the rate-weighted-AP diagram in Figure~\ref{fig:map-rate curve delta qp}. The corresponding BDR values compared to the anchor RDO without delta $\QP$ are listed in Table~\ref{tab: BDR delta qp}. By using the proposed FRDO with SAD, 5.35\,\% of bitrate can be saved for the same detection accuracy of the Mask R-CNN compared to the standard VTM. Contrary, to obtain the same PSNR as the anchor, the bitrate has to be increased by 4.55\,\%, which again indicates that the proposed FRDO preserves, as desired, the detection rate rather than the pixel-fidelity. When allowing delta $\QP$, the hybrid FRDO with SAD even further increases the coding gains for the VCM scenario up to 9.95\,\%. In contrast, allowing delta $\QP$ for conventional RDO only results in 1.15\,\% rate savings for the VCM scenario, which proves that the coding gains are mainly achieved by the FRDO and not only by allowing delta $\QP$.
\begin{table}[t]
	\normalsize
	\vspace{0.5mm}
	\caption{BDR with respect to the particular quality metric when allowing a delta $\QP$ between -3 and 3 using VTM with conventional RDO without delta $\QP$ as anchor for $\QP\in\{12, 17, 22, 27\}$.}
	\label{tab: BDR delta qp}
	\centering
	\begin{tabular}{lrr}
		\hline
		                    & \multicolumn{2}{c}{Quality metric} \\
		Optimization method &  PSNR               & Weighted AP \\  \hline
		RDO	with delta $\QP$		& -0.36\,\% & -1.15\,\%\\
		FRDO-SAD with delta $\QP$	& 4.55\,\%	& -5.35\,\%\\
		HFRDO-SAD with delta $\QP$	& 1.01\,\% & -9.95\,\%\\
		\hline
	\end{tabular}
		\vspace{-4mm}
\end{table}
%\begin{figure}[t]{}
%	\centering
%	
%	
%	\begin{subfigure}[t]{\imSize}
%		\centering
%		\includegraphics[width=\textwidth]{visual_comparison/yuvview_rdo.png}
%	\end{subfigure}
%	\hfil
%	\begin{subfigure}[t]{\imSize}
%		\centering
%		\includegraphics[width=\textwidth]{visual_comparison/yuvview_frdo.png}
%	\end{subfigure}
%	
%	\begin{subfigure}[t]{\imSize}
%		\centering
%		\includegraphics[width=\textwidth]{visual_comparison/yuvview_rdo_blocks.png}
%	\end{subfigure}
%	\hfil
%	\begin{subfigure}[t]{\imSize}
%		\centering
%		\includegraphics[width=\textwidth]{visual_comparison/yuvview_frdo_blocks.png}
%	\end{subfigure}
%	
%	\caption{Comparison of one decoded CTU out of Cityscapes frame \textit{frankfurt\_000000\_000294\_leftImg8bit.png} compressed with RDO (left) and FRDO with SAD (right) and a QP of 22. The corresponding block structure is shown below, where warmer color represents a higher depth in the recursion tree.}
%	\label{fig:visual comparison}
%\end{figure}

\section{Conclusion}

In this paper, a novel feature-based RDO for VCM scenarios has been presented that returns a standard-compliant bitstream optimized for neural networks. With this FRDO, the encoder aims to achieve a high feature-fidelity rather than a high pixel-fidelity. This results in bitrate savings when the detection accuracy of a neural segmentation network is taken to evaluate the coding quality instead of metrics trying to represent the human visual system. In a VCM scenario with the Mask R-CNN network as final detector, HFRDO is able to save up to 5.49\,\% bitrate compared to the standard RDO in VTM-8.0. A drawback in this case is the increased encoder runtime of around 2.3, when encoding with FRDO compared to standard VTM.  Allowing the encoder with HFRDO-SAD to vary the $\QP$ for each CU, further reduces the required bitrate up to 9.95\,\%. Using the VGG-16 network to obtain the feature space that differs from the final evaluation network in structure and weights, indicates the universal applicability of FRDO. Nevertheless, it would be interesting in future work to create the feature space for FRDO with the first few layers of the evaluation network. This can serve as oracle test to see whether the coding performance can be even further increased, when the encoder is adapted to the evaluation network. Besides, FRDO could also be applied in other encoder parts and it could also be tested for inter-coding with a suitable dataset.

\bibliographystyle{IEEEtran}
\bibliography{/home/fischer/Paper/jabref_literature_research_ms2.bib,/home/fischer/Paper/literature_M2M_communication.bib}

\end{document}